\documentclass{article}
\usepackage{spconf,amsmath,graphicx,hyperref}

\title{ROBUST PROVABLY SECURE IMAGE STEGANOGRAPHY \\ VIA LATENT ITERATIVE OPTIMIZATION}

\name{Yanan Li$^{2}$, 
	Zixuan Wang$^{2}$, 
	Qiyang Xiao$^{2}$,
	Yanzhen Ren$^{1,2,*}$%
	\thanks{*Corresponding author: Yanzhen Ren (renyz@whu.edu.cn).}%
	\thanks{This work is supported by the \emph{National Natural Science Foundation of China (NSFC)} under Grant No.~62572358, 62172306, and 62372334.}
}

\address{
	$^{1}$Key Laboratory of Aerospace Information Security and Trusted Computing, Ministry of Education,\\
	$^{2}$School of Cyber Science and Engineering, Wuhan University}

\begin{document}
%
\maketitle
\begin{abstract}
We propose a robust and provably secure image steganography framework based on latent-space iterative optimization. Within this framework, the receiver treats the transmitted image as a fixed reference and iteratively refines a latent variable to minimize the reconstruction error, thereby improving message extraction accuracy. 
Unlike prior methods, our approach preserves the provable security of the embedding while markedly enhancing robustness under various compression and image processing scenarios. 
On benchmark datasets, the experimental results demonstrate that the proposed iterative optimization not only improves robustness against image compression while preserving provable security, but can also be applied as an independent module to further reinforce robustness in other provably secure steganographic schemes. This highlights the practicality and promise of latent-space optimization for building reliable, robust, and secure steganographic systems.

\end{abstract}
\section{Introduction}
\label{sec:intro}

Steganography~\cite{simmons1984prisoners} aims to conceal secret messages within digital media while maintaining the imperceptibility of the embedding process. With the rapid advancement of generative artificial intelligence~\cite{song2020denoising, lu2022dpm, rombach2022high, lu2023dpmsolverfastsolverguided}, provably secure steganographic schemes have attracted increasing attention~\cite{li2025coas,hu2024establishing, jois2024pulsar}. Such schemes guarantee that the distribution of stego objects remains statistically indistinguishable from that of cover objects~\cite{cachin1998information, cryptoeprint:2002/137}. Supported by their rigorous theoretical foundations, provably secure steganography offers strong security assurances, making it particularly valuable in practical applications and adversarial environments~\cite{ding2023discop, wang2025sparsamp, li2025coas,peng2023stegaddpm,bai2025provably}.

However, in practical scenarios, message extraction in steganography often encounters two major challenges. First, transmitted carriers typically undergo lossy operations such as image compression and format conversion~\cite{peng2024ldstega}. Since these operations are inherently nonlinear, they inevitably remove significant amounts of information from the carrier, which in turn causes message loss. This remains a principal obstacle in practice. Second, the extraction process usually involves neural networks, where floating-point computations introduce rounding errors due to limited numerical precision. These two challenges together result in a notable reduction in extraction accuracy for existing provably secure steganographic algorithms, thereby constraining their robustness and practical applicability.

To address these challenges, we propose a latent-space iterative correction strategy inspired by the fixed-point iteration principle~\cite{creswell2018inverting, zhu2024domain}. Specifically, during the extraction phase, we employ a correction mechanism that iteratively adjusts the latent variable, effectively mitigating distortions caused by compression or format transformations. The key idea is to exploit the backpropagation capability of neural networks to iteratively refine the latent variable, guiding it toward the optimal value (i.e., the original message-carrying latent variable). Importantly, this refinement process does not alter the original embedding logic, thus preserving the provable security of the underlying framework. As a result, the proposed fixed-point guided correction substantially improves extraction accuracy across diverse compression conditions while upholding rigorous security guarantees.

In summary, we propose a latent-variable iterative optimization strategy for generative steganography that requires neither modifying the underlying neural framework nor updating its parameters. The strategy is applied solely at the extraction stage, thereby strictly preserving the provable security of the base framework. Extensive experiments demonstrate that the proposed method enhances robustness against image compression and format conversion, while significantly improving message extraction accuracy. 

The independence of the proposed optimization algorithm allows it to be applied to other provably secure steganographic schemes. Experimental results demonstrate that our approach further enhances the robustness of the current SOTA provably secure steganographic model~\cite{hu2024establishing}. These findings underscore both the practicality and the promise of the latent variable iterative optimization strategy.

\begin{figure*}[t]
	\centering
	\includegraphics[width=0.79\textwidth]{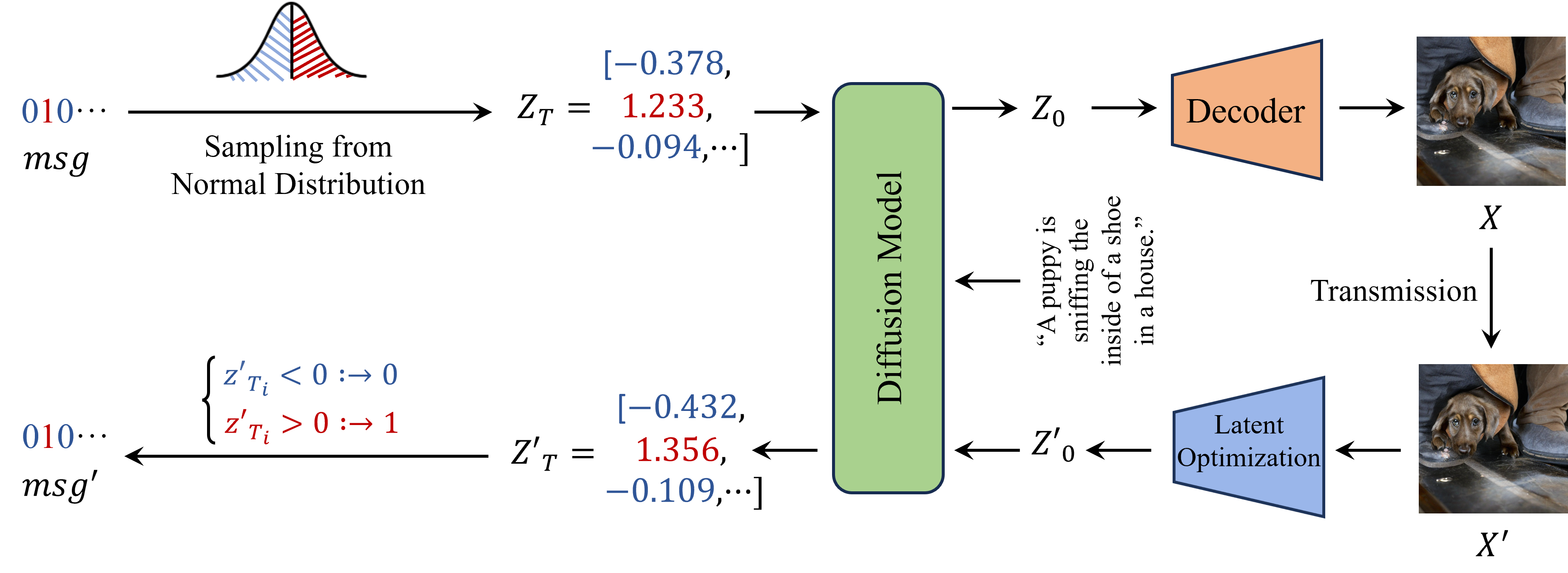}
	\caption{Overall framework of robust and provably secure image steganography via latent-space iterative optimization.}
	\label{fig:frame}
\end{figure*}

\section{Proposed Method}
\label{sec:pagestyle}

\subsection{Embedding Procedure}

As illustrated in Fig.~\ref{fig:frame}, we construct a steganography framework based on latent-space diffusion models. The framework comprises the message $M$, the initial noise variable $Z_T$, intermediate latent variables $Z_i$ during the denoising process, the final latent variable $Z_0$, and the decoded stego image $X$. During transmission, the image $X$ may undergo compression or format conversion, resulting in a distorted received image $X'$. Correspondingly, a series of latent variables $Z_i'$ can be obtained during the inversion process.

Assume the message to be transmitted is $M=\{0,1\}^n$, which has been encrypted. For simplicity, we assume that the bits in $M$ follow an independent Bernoulli distribution with parameter $p=0.5$. 
\[
M = \{m_i\}_{i=1}^n, \quad m_i \overset{\text{i.i.d.}}{\sim} \text{Bernoulli}(0.5).
\]

Specifically, for any bit $m_i \in M$:
\begin{itemize}
	\item If $m_i=0$, a value $s_i$ is uniformly sampled from the interval $(0,0.5)$;
	\item If $m_i=1$, a value $s_i$ is uniformly sampled from the interval $[0.5,1)$.
\end{itemize}

After sampling all bits, we obtain the set $S=\{s_i\}_{i=1}^n$, which follows a uniform distribution. Using the probability integral transform of the Gaussian distribution, $S$ is then mapped to a high-dimensional latent variable $Z_T$, which follows the standard Gaussian distribution:
\begin{equation}
	z_{T_i} = \Phi^{-1}(s_i)
	\label{eq:latent_init}
\end{equation}
where $\Phi^{-1}(\cdot)$ denotes the inverse cumulative distribution function of the standard Gaussian distribution, $z_{T_i} \in Z_T$. Through the denoising process of the diffusion model, $Z_0$ is obtained, which is subsequently decoded to generate the stego image $X$.

\subsection{Extraction Procedure}

At the receiver, the latent variable sequence is first encoded and then iteratively optimized, yielding the recovered latent variables $Z_T'=\{z_{T_i}'\}_{i=1}^n$. Based on Eq.~\ref{eq:latent_init}, it follows that:
\begin{align*}
	s_i \in (0,0.5)  &\ \Longleftrightarrow\  z_{T_i} < 0 \\
	s_i \in [0.5,1)  &\ \Longleftrightarrow\  z_{T_i} \ge 0
\end{align*}

Based on this relationship, a simple zero-threshold decision rule is applied for decoding:
\[
\hat{m}_i =
\begin{cases}
	0, & z_{T_i}' < 0\\[2mm]
	1, & z_{T_i}' \ge 0
\end{cases}
\]
That is, when the recovered latent variable $z_{T_i}'$ is less than 0, the corresponding $s_i$ falls in $(0,0.5)$ and is decoded as bit 0; otherwise, it is decoded as bit 1.

\subsection{Latent Optimization}
\begin{figure}[!t]
	\centering
	\includegraphics[width=0.75\columnwidth]{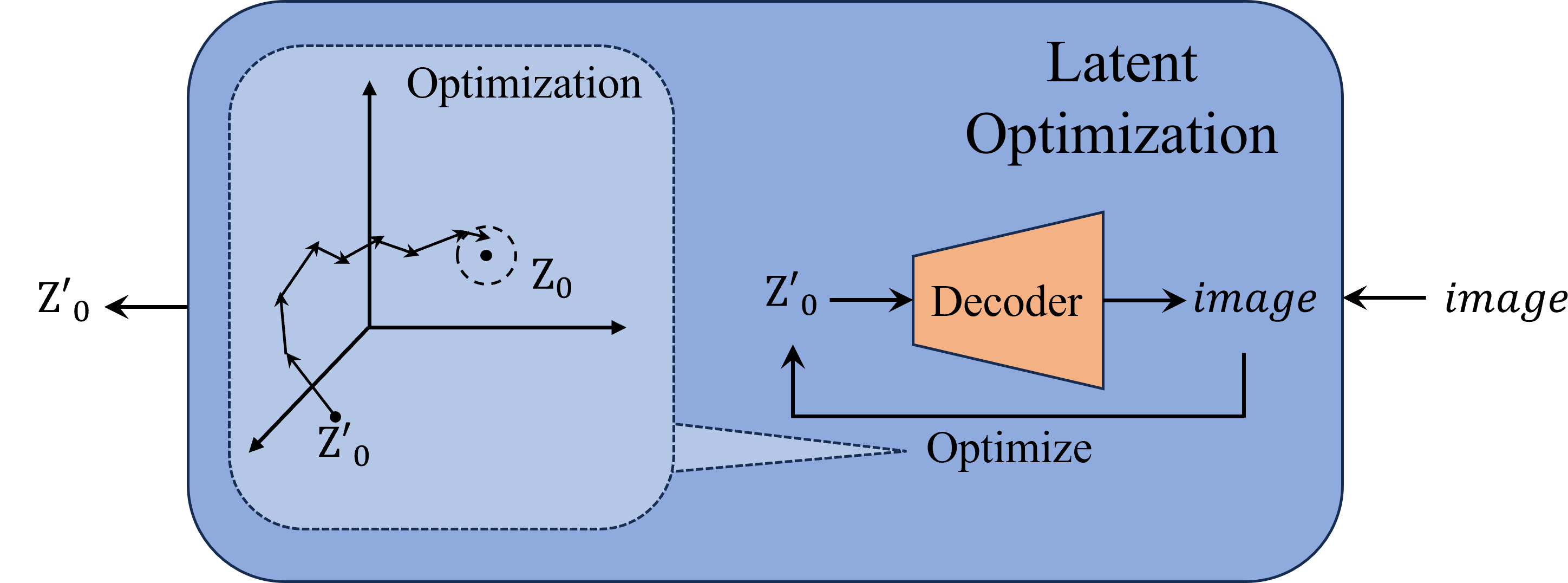}
	\caption{Illustration of latent-space optimization and the trajectory of latent variables.}
	\label{fig:latent_opt}
\end{figure}

In steganographic communication, the transmitted image $X$ is perturbed by the channel and received as $X'$, which serves as the only invariant at the decoder. The primary objective is to refine a latent variable $Z'_{0_i}$ such that the decoded image $D(Z'_{0_i})$ matches $X'$, thereby improving extraction accuracy. To achieve this, we adopt an iterative latent optimization scheme inspired by fixed-point methods, as shown in Fig.~\ref{fig:latent_opt}.

The initial latent variable $Z'_{0_1}$ is obtained from the encoder:
\begin{equation}
	Z'_{0_1} = E(X')
\end{equation}
and is subsequently refined through iterative optimization, where the index $i$ in $Z'_{0_i}$ denotes the iteration step.

We define the reconstruction loss as:
\begin{equation}
	\mathcal{L}(Z'_{0_i}) = \ell(D(Z'_{0_i}), X') \qquad
	\ell(\hat{X}, X') = \frac{1}{2}\|\hat{X}-X'\|_2^2
\end{equation}

where $D$ denotes the decoder in the diffusion model. The iterative latent update is then given by:
\begin{equation}
	Z'_{0_{i+1}} = f_{\eta}(Z'_{0_i}) = Z'_{0_i} - \eta \nabla \mathcal{L}(Z'_{0_i})
	\label{eq:bound0}
\end{equation}
where $\eta$ is the step size. In this work, we set $\eta = 1$.

For any $Z'_{0_i}, Z'_{0_{i+1}}$, according to Eq.~\ref{eq:bound0}, we have:
\begin{equation}
	\|Z'_{0_{i+1}} - Z'_{0_i}\| = \eta \, \|\nabla \mathcal{L}(Z'_{0_i})\|
	\label{eq:bound1}
\end{equation}

By the chain rule, for the squared $\ell_2$ loss, we have $\nabla \mathcal{L}(Z'_{0_i}) = J_D(Z'_{0_i})^\top (D(Z'_{0_i})-X')$.  
Assume $D$ is differentiable and its Jacobian is bounded as $\|J_D(Z)\|_2 \le L_J$ (e.g., $D$ is $L_J$-Lipschitz~\cite{kinoshita2023controlling}). Then:
\begin{equation}
	\|\nabla \mathcal{L}(Z'_{0_i})\| \le L_J \|D(Z'_{0_i})-X'\|
	\label{eq:Lipschitz}
\end{equation}

By substituting the Lipschitz condition~\ref{eq:Lipschitz} into inequality~\ref{eq:bound1}, we obtain:

\begin{equation}
	\|Z'_{0_{i+1}} - Z'_{0_i}\| 
	\le \eta L_J \|D(Z'_{0_i})-X'\|
	\label{eq:bound2}
\end{equation}

If the reconstruction error decreases such that:
\begin{equation}
	\|D(Z'_{0_i})-X'\|\;\to\; 0
	\label{eq:reconvergence}
\end{equation}
then Eq.~\ref{eq:bound2} implies that $\|Z'_{0_{i+1}} - Z'_{0_i}\| \to 0$.

That is, the update magnitude diminishes as $D(Z'_{0_i})$ approaches $X'$, and the generated outputs progressively approximate $X'$, leading to improved message extraction accuracy in practice.

In other words, since $X'$ is fixed during decoding, we minimize:
\[
\min_{Z'_0} \|D(Z'_0)-X'\|_2^2
\]
to reduce the reconstruction discrepancy. This gradient-based refinement drives the reconstruction toward $X'$ and typically improves extraction accuracy, even without assuming strict global convergence guarantees.

\subsection{Provable Security Analysis under Latent-Space Optimization}

In our latent-space diffusion-based steganography framework, the embedding process maps the encrypted message $M=\{0,1\}^n$ to latent variables $Z_t$ via a two-step procedure. First, each message bit $m_i$ is mapped to a uniform sample $s_i$:
\begin{itemize}
	\item if $m_i = 0$, then $s_i \sim \mathrm{Uniform}(0,0.5)$;
	\item if $m_i = 1$, then $s_i \sim \mathrm{Uniform}[0.5,1)$.
\end{itemize}

Collecting all $s_i$ gives $S = \{s_i\}_{i=1}^n$, which is strictly uniformly distributed over $(0,1)$. Then, $S$ is mapped to the latent space via the inverse CDF of the standard Gaussian distribution (probability integral transform):
\begin{equation*}
	Z_t = \Phi^{-1}(S)
\end{equation*}
Consequently, $Z_t$ follows exactly the standard Gaussian distribution, identical to the initial noise distribution used in the diffusion model. 

Since the latent variable $Z_t$ is statistically indistinguishable from the standard Gaussian noise, the KL divergence between the stego latent distribution $p_{Z_t}$ and the standard Gaussian distribution $p_{\mathcal{N}}$ satisfies:
\begin{equation*}
	D_{\mathrm{KL}}(p_{Z_t} \,\|\, p_{\mathcal{N}}) = 0
\end{equation*}

Importantly, the iterative latent-space optimization at the receiver is entirely independent of the embedding process. Let $Z_0'$ denote the latent variable recovered from the received image $X'$, which is refined via gradient-based optimization to improve message extraction accuracy. Since the embedding distribution $p_{Z_t}$ and the corresponding stego image distribution $p_X$ are unchanged by this receiver-side optimization, the provable security of the scheme is fully preserved~\cite{cachin1998information}.

In summary, the provable security derives directly from the uniform-to-Gaussian mapping of the encrypted message, and the proposed latent-space optimization at the receiver enhances extraction accuracy without compromising security.

\section{Experiments}
\label{sec:typestyle}

\subsection{Setup}
All experiments were conducted on a workstation equipped with a single NVIDIA RTX 4090 GPU. The image latent-space diffusion model is based on the open-source Stable Diffusion 2.1~\cite{huggingface_diffusers, lu2023dpmsolverfastsolverguided}, with a latent-space size of $4 \times 64 \times 64$, which generates images at a fixed resolution of $512 \times 512$. The generation prompts are primarily drawn from the COCO dataset~\cite{lin2014microsoft}. 

In our experiments, we traded off embedding capacity for enhanced robustness, resulting in an effective capacity of $0.0625 = \frac{4 \times 64 \times 64}{512 \times 512}$. We evaluated robustness under various image compression formats, including lossless TIFF, PNG, and three levels of JPEG compression (Q90, Q70, and Q50, where lower quality corresponds to higher compression).

\subsection{Performance}

\begin{table}[t]
	\centering
	\caption{Message extraction accuracy under different compression formats.}
	\label{tab:compression_results1}
	\resizebox{\columnwidth}{!}{%
		\begin{tabular}{c|cccccc}
			\hline
			Method & TIFF32 & TIFF16 & PNG & JPEG90 & JPEG70 & JPEG50 \\
			\hline
			Hu & 0.9830 & 0.9820 & 0.9805 & 0.9634 & 0.9235 & \textbf{0.8887} \\
			Ours & 0.9354 & 0.9349 & 0.9288 & 0.9118 & 0.8776 & 0.8421 \\
			Ours (Opt) & \textbf{0.9877} & \textbf{0.9876} & \textbf{0.9854} & \textbf{0.9683} & \textbf{0.9272} & 0.8820 \\ 
			\hline
		\end{tabular}
	}
\end{table}

Table~\ref{tab:compression_results1} reports the message extraction accuracy under different image compression formats for three methods: the original Hu~\cite{hu2024establishing} framework, our baseline method (``Ours''), and our optimized version (``Ours (Opt)''). For both Hu's method and ours, the extraction accuracy decreases noticeably as the compression level increases, from TIFF32 to JPEG50. With the proposed optimization, our method consistently improves accuracy across all formats. Compared with Hu's original method, the optimized version achieves comparable performance in most cases and demonstrates clear advantages on all formats except JPEG50. These results demonstrate that the proposed optimization effectively reinforces message recovery and enhances the practical robustness of provably secure steganography across diverse image formats.

\subsection{Ablation}

\begin{figure}[!t]
	\centering
	\includegraphics[width=0.70\columnwidth]{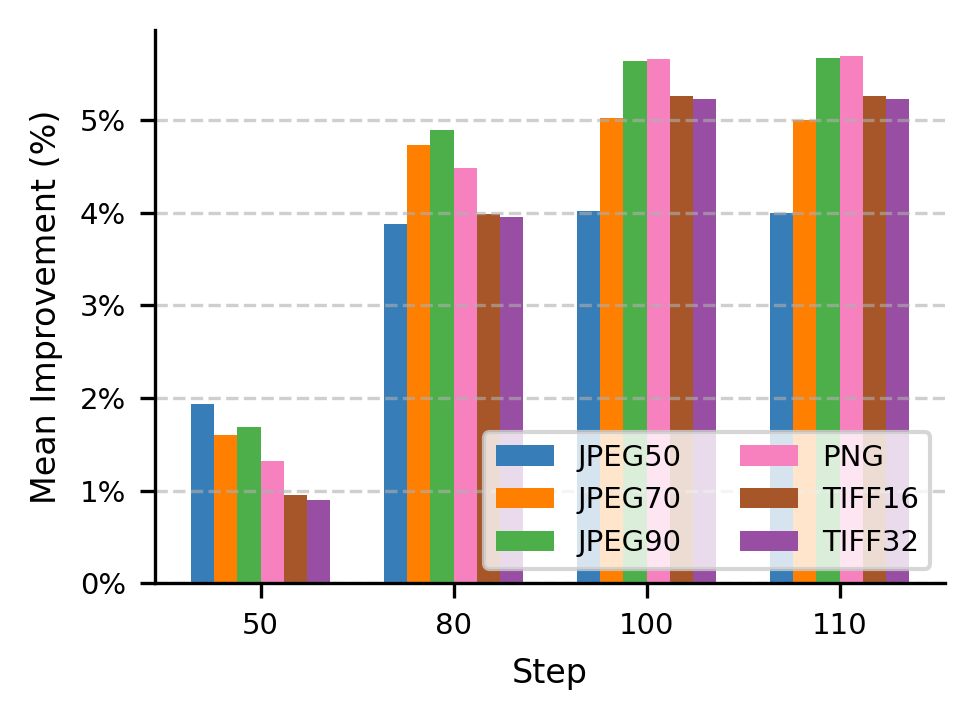}
	\caption{Mean extraction accuracy gain (\%) for six image formats at 50/80/100/110 latent-space optimization steps.}
	\label{fig:stepvsacc}
\end{figure}

Fig.~\ref{fig:stepvsacc} presents the mean improvement (\%) of message extraction accuracy for six image formats under four different optimization steps (50, 80, 100, 110). As observed, all formats exhibit a monotonic increase in improvement as the number of optimization steps increases from 50 to
100, indicating that the iterative optimization effectively enhances extraction performance.

For example, the JPEG90 format shows an improvement of 1.68\% at step 50, which increases to 5.64\% at step 100 and remains stable at 5.67\% at step 110. Similar trends are observed for PNG (from 1.32\% to 5.67\%), TIFF16 (from 0.95\% to 5.27\%), and TIFF32 (from 0.9\% to 5.23\%). These results suggest that the proposed optimization consistently benefits both lossy and lossless formats, with slightly higher gains observed for high-quality lossy formats (JPEG90, JPEG70) and PNG compared to low-quality lossy formats (JPEG50).

Interestingly, for all formats, the mean improvement saturates around step 100--110, indicating diminishing returns beyond 100 steps. 

Overall, these results confirm that the proposed iterative optimization strategy substantially enhances message extraction accuracy, particularly under high-capacity embedding conditions, while maintaining consistent performance across diverse image formats.

\subsection{Cross-Model Applicability of Latent-Space Optimization}

\begin{figure}[!t]
	\centering
	\includegraphics[width=0.70\columnwidth]{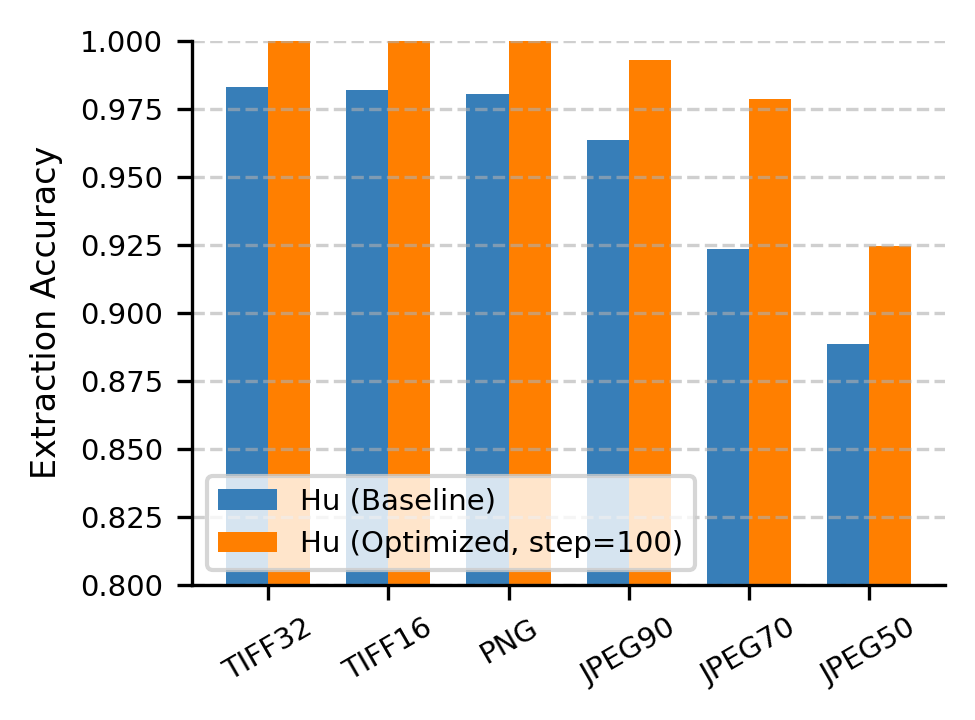}
	\caption{Generality of latent-space iterative optimization: Hu accuracy across formats (baseline vs. 100 steps).}
	\label{fig:hu}
\end{figure}

To evaluate the practical effectiveness of the proposed optimization scheme, we applied it to Hu's steganographic framework~\cite{hu2024establishing}. Figure~\ref{fig:hu} presents a grouped bar chart comparing message extraction accuracy across six image formats before and after optimization. The baseline extraction accuracy ranges from 0.9830 for TIFF32 to 0.8887 for JPEG50. After applying the optimization (step = 100), the accuracy for lossless formats improves to nearly perfect levels, while lossy formats also exhibit notable gains, e.g., 0.9855 for JPEG70 and 0.9285 for JPEG50. These results clearly demonstrate that the optimization significantly enhances extraction accuracy across both lossless and lossy formats, validating its practical utility in strengthening the robustness of provably secure steganographic schemes.

\section{Conclusion}
\label{sec:majhead}


This paper proposes a latent iterative optimization scheme that is executed at the receiver. During the extraction phase, it iteratively refines the latent variables, improving message recovery accuracy under various compression and format conversion scenarios. The optimization process does not compromise the security of the original framework, making it applicable to other similar models. Essentially, this approach trades additional time and computational resources for higher accuracy, which is entirely reasonable and acceptable in steganographic tasks where security is the primary concern.


\vfill\pagebreak

\bibliographystyle{IEEEbib}
\bibliography{strings,refs}

\end{document}